\documentclass[12pt]{article}
\usepackage[margin=2cm]{geometry}
\usepackage{comment}
\usepackage{amsmath,amssymb,extarrows,graphicx,subfigure,setspace}
\usepackage{cite}
\usepackage{slashed}
\usepackage{color}
\makeatother

\usepackage{authblk}
\usepackage{color}
\usepackage{array}
\usepackage[utf8]{inputenc}

\usepackage{tensor}

\usepackage{fullpage}
\usepackage{url}
\usepackage{tikz}
\usetikzlibrary[topaths]

\usetikzlibrary{lindenmayersystems}
\usetikzlibrary[shadings]
\usepackage{pgfplots}

\usepackage{amsthm}
\usepackage{amsmath}
\usepackage{amssymb}

\usepackage{mathtools}

\DeclareMathOperator{\Tr}{Tr}

\usepackage[subnum]{cases}

\usepackage{mathrsfs}

\newcommand{\bse}{\begin{subequations}}
\newcommand{\ese}{\end{subequations}}
\newcommand{\be}{\begin{equation}}
\newcommand{\ee}{\end{equation}}
\newcommand{\bea}{\begin{eqnarray}}
\newcommand{\eea}{\end{eqnarray}}
\newcommand{\ba}{\begin{array}}
\newcommand{\ea}{\end{array}}
\newcommand{\nn}{{\nonumber}}

\newcommand{\KK}{\mathcal{K}}

\usepackage{amssymb}

\theoremstyle{definition}

\theoremstyle{remark}

\begin{document}
\onehalfspacing
\noindent

\begin{titlepage}
\vspace{10mm}
\begin{flushright}
 %IPM/P-2015/nnn \\
%FPAUO-1/10\\
\end{flushright}

\vspace*{20mm}
\begin{center}

{\Large {\bf On Butterfly effect in Higher Derivative  Gravities}\\
}

\vspace*{15mm}
\vspace*{1mm}
{Mohsen Alishahiha${}^a$,   Ali Davody$^{b}$,  Ali Naseh${}^b$ and   Seyed Farid Taghavi${}^{b}$}

 \vspace*{1cm}

{\it ${}^a$ School of Physics, ${}^b$  School of Particles and Accelerators,\\
Institute for Research in Fundamental Sciences (IPM)\\
P.O. Box 19395-5531, Tehran, Iran  }

 \vspace*{0.5cm}
{E-mails: {\tt alishah,davody,naseh,s.f.taghavi@ipm.ir}}%

\vspace*{1cm}
%%\maketitle
\end{center}

\begin{abstract}
We study butterfly effect in $D$-dimensional  gravitational theories containing terms 
quadratic in Ricci scalar and Ricci tensor.  One  observes  that due to higher order 
derivatives in the corresponding equations of motion there are two butterfly velocities. 
The velocities are determined by the dimension of operators 
whose sources are provided by the metric.  The three dimensional TMG model is also studied 
where we get two butterfly velocities at generic point of the moduli space of parameters. 
At critical point two velocities coincide.
\end{abstract}

\end{titlepage}

\section{Introduction}
In the context of  gauge/gravity duality a thermal system may be described by a gravitational 
theory on an AdS black hole solution\cite{Witten:1998zw} which can be used to explore different 
aspects of a thermal system such as chaos.  Indeed it was shown \cite{{Shenker:2013pqa},{Shenker:2013yza},{Leichenauer:2014nxa},{Roberts:2014isa}} that
chaos in a thermal CFT may be described by shock waves near the horizon of an AdS black hole. More precisely, the black hole geometry may be perturbed by a small 
perturbation and due to the back-reaction the perturbation  grows in time resulting 
to a  geometry which is given by a shock wave propagating on the horizon of the black hole.
In other words, holographically  the propagation of the shock wave on the horizon 
would provide a description of butterfly effect  in the dual field theory. 

On the other hand in the field theory side the butterfly effect may be diagnosed  by out-of-time order four-point function between pairs of local operators
\be
\langle V_x(0) W_y(t) V_x(0) W_y(t)\rangle_\beta\,
\ee
where $\beta$  indicates a thermal expectation value. In terms of  this correlation function, 
the butterfly effect may be seen by a sudden decay after the {\it scrambling time}, $t_*$, 
\be
\frac{\langle V_x(0) W_y(t) V_x(0) W_y(t)\rangle_\beta}
{\langle V_x(0) V_x(0)\rangle_\beta\langle W_y(t)  W_y(t)\rangle_\beta}\sim 1- e^{\lambda_L\left(t-t_*+
\frac{|x-y|}{v_B}\right)},
\ee
where $\lambda_L$ is the Lyapunov exponent and $v_B$ is butterfly velocity. From gravity 
point of view, this four-point function can be holographically computed from the certain 
component of the back-reacted metric \cite{Shenker:2013pqa} and thus the butterfly velocity 
should be identified with the velocity of shock wave by 
which the perturbation spreads in the space. The Lyapunov exponent is given in terms
of the Hawking temperature, $\lambda_{L}=\frac{2\pi}{\beta}$.
 
 The aim of this paper is to further study butterfly effect in gravitational theories containing 
 higher derivative terms. We note, however, that butterfly effect for Gauss-Bonnet action and 
 an action containing quadratic terms have been partially studied in literature (see {\it e.g.}
 \cite{{Roberts:2014isa},{Mezei:2016wfz},Sircar:2016old}). In the present paper we would like to extend these works
 in more details. In particular, we will show that for theories whose gravitational dual are  provided by the Einstein gravity modified by terms quadratic in Ricci
 scalar and  Ricci tensor, one generally finds two butterfly velocities which are given by 
 graviton excitations on the boundary. This is also the case for three dimensional TMG model.
 
Actually this is a generic feature of higher derivative gravity whose equations of motion are
higher order differential equations. The precise number of butterfly velocities are given by the
number of boundary conditions needed to fix the metric.

This may be understood as follows: Indeed it was shown\cite{Perlmutter:2016pkf}
that in any holographic CFTs whose gravitational description is provided by the Einstein 
gravity, the butterfly velocity is determined by the spin-2 operator of lowest twist that is the
energy-momentum tensor of dual boundary theory. On the other hand, from holographic renormalization\cite{deHaro:2000xn} it is known that the boundary value of the metric provides a source for the energy-momentum tensor. When the action contains higher derivative terms, generally the corresponding equations of motion are higher order differential equations. Therefore to fix the metric one needs more than 
one boundary value. This, in turn, indicates that boundary values of metric provide sources for 
more than one operator. 

On the other hand since, generically, by tuning the parameters of the model one can 
make  the dimensions of these extra operators as closed as that  of energy-momentum tensor, their contributions could be as important as the energy-momentum tensor. Therefore for each of these operators one has  a butterfly velocity that is given in terms of its  dimension.

The paper is organized as follows: In the next section, we study shock wave in 
$D$-dimensional  gravity where we find that at a genetic point there are two butterfly 
velocities, while at the critical point two velocities coincide. In section three, we will redo 
the same computations for TMG model where, again, we get two butterfly velocities. 
In this case, we will also reproduce the resultant velocities from the dual 2D conformal field theory 
where one shows that the butterfly velocity is given in terms of dimension of  operators dual 
to the perturbation of  metric. The last section is devoted to conclusions. 
%%%%%%%%%%%%%%%%%%%%%%%%%%%%%%%%%%%%%%%%%%
%%%%%%%%%%%%%%%%%%%%%%%%%%%%%%%%%%%%%%%%%%%
\section{Shock wave in higher derivative gravity}
In this section, we would like to study  butterfly effect in $D$-dimensional gravitational theories 
consisting of Einstein gravity modified by  certain $R$-squared terms. The action we will be 
considering is
\be
I=\frac{1}{\kappa}\int d^{D}x\,\sqrt{-g}\left[R+\frac{(D-1)(D-2)}{\ell_0^2}+{\alpha}_1 R^{2}+\alpha_2 
R^{\mu\nu}R_{\mu\nu}
\right],\label{eq:Quadratic_action}
\ee
where $\alpha_1$ and $\alpha_2$ are free parameters, $\ell_0$ is a length scale. This model, 
has been studied in the literature 
(see {\it.e.g.}\cite{{Lu:2011zk},{Alishahiha:2011yb},{Deser:2011xc}}) where it was 
shown that at a generic point of moduli space of  parameters, excitations above an AdS vacuum 
contain scalar ghost, massive and massless spin-2 gravitons. Nevertheless, it is possible 
to remove the scalar ghost by tuning  parameters $\alpha_1$ and $\alpha_2$
\cite{{Stelle:1976gc},{Stelle:1977ry}}. Moreover at {\it critical points}, the massive
spin-2 becomes massless leading to a logarithmic mode\cite{Alishahiha:2011yb}. 
 
The corresponding equations of motion are given by $\mathcal{E}_{\mu\nu}=0$ 
with\cite{Gullu:2009vy}
\bea\label{higherEQs}
&&\mathcal{E}_{\mu\nu}= R_{\mu\nu}-\frac{1}{2}g_{\mu\nu}R -\frac{(D-1)(D-2)}{2\ell_0^2}
 g_{\mu\nu}
+2{\alpha}_1\left(R_{\mu\nu} -\frac{1}{4}g_{\mu\nu}R+g_{\mu\nu}\square-\nabla_{\mu}\nabla_{\nu}
\right)R\cr &&
+\hspace{.5mm}\alpha_2\left[\left(g_{\mu\nu}\square-\nabla_{\mu}\nabla_{\nu}\right)R+
\square\left(R_{\mu\nu}-\frac{1}{2}g_{\mu\nu}R\right)+2\left(R_{\mu\sigma\nu\rho}
-\frac{1}{4}g_{\mu\nu}R_{\sigma\rho}\right)R^{\sigma\rho}\right].
\label{fieldequations}
\eea
For generic values of the parameters $\ell_0$, ${\alpha}_1$ and $\alpha_2$  the model has two 
distinct vacua such that $R_{\mu\nu}=\frac{D-1}{\ell^2}g_{\mu\nu}$, with $\ell$ being a
root of the following equation\cite{Deser:2011xc}\footnote{In four dimensions or for the case 
of $D{\alpha}_1+\alpha_2=0$ the equation has a single solution $\ell^2=\ell_0^2$. Note that 
for generic values of parameters, it is always possible
to tune the parameters such that at least one of the vacua to be an
AdS$_D$  geometry.}
\be\label{Lam}
\ell^2(\ell^2-\ell_0^2)+ \frac{(D-4)(D-1)}{D-2}(D{\alpha}_1+\alpha_2)\ell_0^2=0.
\ee
Then it is straightforward to show that the above equations of motion admit asymptotically
AdS black brane solution   
 \be\label{SchwarzMetric}
ds^2=-f(r)dt^2+f^{-1}(r)dr^2+\frac{r^2}{\ell^2}d\vec{x}^2,\quad f(r)=\frac{r^2}{\ell^2}\left(1-
\frac{r_h^{D-1}}{r^{D-1}}\right),
\ee 
where $r_h$ is the radius of horizon.

Now the aim is to study a shock wave solution of this model when 
the above black hole solution is perturbed by  injecting a small amount  of energy.
To proceed, it is useful to re-write the black brane solution in the Kruskal coordinates
\be
u=\exp[\frac{2\pi}{\beta}\left(r_*-t\right)],\qquad v=-\exp[\frac{2\pi}{\beta}\left(r_*+t\right)],
\ee
where $\beta=4\pi/f'(r_h)$ is the inverse of the temperature and $dr_*=dr/f(r)$ is the 
tortoise coordinate whose near horizon expression is 
\bea
r_*&=&\frac{\beta}{4\pi}\bigg{[}\log\frac{r-r_h}{2r_h}-\log c+ \frac{D-4}{2 {r_h}} (r-r_h)
+\frac{(D-14) D+36}{24 {r_h}^2} (r-r_h)^2\cr &&-\frac{(D-6)(5D-16)}{72 {r_h}^3}
(r-r_h)^3+
{\cal O}((r-r_h)^4)\bigg{]},
\eea
and $c$ is a positive number to be fixed latter. 
By making use of this coordinate system, the metric can be 
recast into the following form
\be
ds^2=2 A(uv) dudv+B(uv) d\vec{x}^2.
\ee
Here $A(uv)$ and $B(uv)$ are two functions, implicitly,  given by the component of 
the black brane metric $f$, whose near horizon expansions are 
\bea
&& A(x)=-\frac{4 c \ell^2}{D-1}\bigg(1+2 c (D-4) x+c^2 \left(4 D^2-29 D+54\right) x^2\nn\\
&&\hspace{3cm}+\frac{4}{9} c^3 \left(18 D^3-185 D^2+646 D-768\right) x^3+\cdots\bigg),\nn\\
&& B(x)=\frac{r_h^2}{\ell^2}\bigg( 1-4 c x-4 c^2 (D-5) x^2-\frac{1}{3} 4 c^3 \left(4 D^2-35 D
+78\right) x^3+\cdots \bigg).\nn
\eea
Actually since we are going to study the back-reacted geometry near the horizon, 
the above expressions are sufficient to study the shock wave solution.

Now let us consider an injection of  a small amount of  energy from boundary towards
the horizon at time $-t_w$. This  will cross the $t=0$ time slice while it is red shifted. Therefore 
the equations of motion should be deformed as
\be\label{EOMs}
\mathcal{E}_{\mu\nu}=\kappa T^S_{\mu\nu},
\ee
where the energy-momentum tensor associated with the energy injection which has only $uu$ component is given by 
\be\label{shockEnergy}
T_{uu}^{S}=\ell E e^{2\pi t_w/\beta}\delta(u)\delta^{D-2}(\vec{x}).
\ee
Now the aim is to solve the equations of motion near the horizon to find the shock wave solution.
To proceed, by making use of the  step function $\Theta(x)$  one may consider the following 
ansatz for  the back-reacted geometry 
\be\label{Ans}
ds^2=2A(UV)\,dU\,dV+B(UV) dx^2-2A(UV)h(x)\delta(U)dU^2,
\ee
where  the new coordinates $U$ and $V$ are defined by
\be
U\equiv u,\hspace{1cm} V\equiv v+h(\vec{x})\Theta(u).
\ee
Here $h(x)$ is a function to be found by the equations of motion \eqref{EOMs}. 
Plugging the ansatz \eqref{Ans} into the above equations,  near the horizon at the leading order one
 finds a fourth order differential equation for $h(x)$ 
\bea
&&\left(\frac{\ell^2}{r_h^2}\partial_i\partial^i -\frac{(D-1)(4D{\alpha}_1+(D+2) \alpha_2)-2\ell^2}
{2\alpha_2\ell^2}
 \right)\left(\frac{\ell^2}{r_h^2}\partial_i\partial^i-\frac{(D-1)(D-2)}{2\ell^2}\right)h(x^i)\cr &&\cr &&
 \hspace{7.2cm}
=-\frac{(D-1)}{4\alpha_2}\frac{1}{ c\ell^2}\left[\kappa\,\ell\,E e^{2\pi t_w/\beta}\right]\delta^{D-2}(x^i),
\eea
which can be reduced into two second order differential equations as follows
\begin{numcases}
\displaystyle \left(\partial_i\partial^i -a_1^2 \right)q(x^i)= \eta \,\delta^{D-2}(x^i)\label{DfirstEQ}\\
\displaystyle \left(\partial_i\partial^i-a_2^2\right)h(x^i)=q(x^i),\label{DseqondEQ}
\end{numcases}
where
\bea
&&a_1^2= {\frac{(D-1)(4D{\alpha}_1+(D+2) \alpha_2)-2\ell^2}{2\alpha_2\ell^2}}\;\frac{r_h^2}{\ell^2}
,\qquad a_2^2={\frac{(D-1)(D-2)}{2\ell^2}} \;\frac{r_h^2}{\ell^2},\nn\\
&&\eta=-\frac{(D-1)}{4\alpha_2}\frac{r_h^4}{ c\,\ell^6}\left[\kappa\,\ell\,E e^{2\pi t_w/\beta}\right].
\eea
To simplify the computations, it is useful to use  the symmetry of the background to study 
a shock wave which is a plane-wave propagating in $x=x_1$ direction. This can be done by 
injecting energy along $x$, leading to the energy-momentum $T_{uu}^S=\ell E e^{2\pi t_w/\beta}\delta(u)\delta(x)$ so that the  equation  \eqref{DfirstEQ} reduces to 
 $\left(\partial_i\partial^i -a_1^2 \right)q(x)= \eta \,\delta(x)$ whose solution is
 %  equation  can be solved resulting to
\be\label{GFh}
q(x)=-\frac{\eta}{2a_1}e^{-a_1 |x|},
\ee
%which is, indeed, the  Green's function of the  equations \eqref{DfirstEQ} and \eqref{DseqondEQ}. 
where $|x|$ denotes the absolute value of $x$. From the equation \eqref{DseqondEQ}, it is clear that  $q(x)$ should be thought of as a source for the function $h(x)$. Moreover 
taking into account that the Green's function of the equation \eqref{DseqondEQ} has the same form 
as that of $q(x)$ one finds
\be
h(x)=\frac{\eta}{4a_1 a_2}\int_{-\infty}^{\infty} dy\,e^{-a_1|y|-a_2|x-y|}.
\ee
It is now easy to evaluate this integral to find $h(x)$. To proceed one  assumes  that $x>0$ 
(we get the same result for $x<0$) in which the above expression reads
\bea
h(x)=\frac{\eta}{4a_1a_2} \left[e^{-a_2x}\int_{-\infty}^0 dy\,e^{(a_1+a_2)y} +e^{-a_2x}
\int_{0}^x dy\,e^{-(a_1-a_2)y} 
+ e^{a_2x}\int_x^{\infty} dy\,e^{-(a_1+a_2)y}   \right].
\eea
So that 
\be
h(x)=\frac{\eta}{2 a_1 a_2}\frac{a_1 e^{-a_2 x}-a_2 e^{-a_1 x}}{a_1^2-a_2^2}.
\ee
Using the explicit expressions of $\eta$, $a_1$ and $a_2$ and for an appropriate choice
of $c$, one gets
\be\label{NMGdecay}
h(x)={\frac{\ell^2\sqrt{1/2(D-1)(D-2)}}{\left(\ell^2-2(D-1)(D\alpha_1+\alpha_2)\right)}} \left[v_B^{(1)}\,e^{\frac{2\pi}{{\beta}}
	\left[(t_w-t_*)-|x|/v_B^{(1)}\right]}-v_B^{(2)}\,e^{\frac{2\pi}{{\beta}}\left[(t_w-t_*)-|x|/v_B^{(2)}\right]}
\right],
\ee
where the scrambling time is defined by $t_*=-\frac{\beta}{2\pi}\log \frac{\kappa}{ \ell^{d-2}}$. From this expression one can read  two different butterfly velocities as follows 
\bea\label{BF}
v_B^{(1)}&=&\frac{2\pi}{\beta a_2}=\sqrt{\frac{D-1}{2(D-2)}},\cr &&\cr\;\;\;\;\;
v_B^{(2)}&=&
\frac{2\pi}{\beta a_1}=\sqrt{\frac{D-1}{2(D-2)}}\;
%\frac{a_2}{a_1}.
\sqrt{\frac{(D-1)(D-2)\alpha_2}{(D-1)(4D{\alpha}_1+(D+2)\alpha_2)-2\ell^2}}.
\eea

As we have already mentioned, the model under consideration given by 
the action \eqref{eq:Quadratic_action} above its AdS vacuum has different propagating modes
including  massive and massless spin-2  modes. 
The mass of the massive graviton is also given by
\be
M^2=\frac{2(D-1)(D\alpha_1+\alpha_2)-\ell^2}{\alpha_2\ell^2}.
\ee
It is then interesting to re-write the second  butterfly velocity in the equation  \eqref{BF}  in terms of 
the mass $M^{2}$,
\be
v_B^{(2)}=\sqrt{\frac{D-1}{2(D-2)}}\;\frac{1}{\sqrt{1+\frac{2\ell^2}{(D-1)(D-2)}M^2}}.
\ee
It is then clear that at the critical point where the massive graviton degenerates with the massless
graviton, $M^{2}=0$, two velocities coincide  resulting to one butterfly velocity. In this 
 case the model exhibits a logarithmic mode\cite{Alishahiha:2011yb} and therefore  the expression 
of metric perturbation $h(x)$ gets modified as follows
\be \label{logNMG}
h(x)=\left(\frac{\ell^2 v_B(v_B+2\pi|x|/\beta)}{-(D-1)^2(D-2)\alpha_2}\right) e^{\frac{2\pi}
	{{\beta}}\left[(t_w-t_*)-|x|/v_B\right]},
\ee
where $v_B=v^{(1)}_B$ is the butterfly velocity at the critical point. 
 
These results may be understood as follows. Actually in the context of AdS/CFT correspondence 
there is a correspondence between bulk fields and boundary operators in the sense that 
the boundary value of the bulk field provides a source for the boundary operator. In particular 
the energy-momentum tensor in the boundary theory is sourced by the metric. We note, however,
that when the equations of motion of a bulk field contains higher order derivatives, 
the corresponding field could provide sources for different operators on the boundary.
 
In the present case where we are dealing with higher derivative gravity the equations of motion
of the metric are fourth order and the metric provides two sources for two operators corresponding 
to massless and massive gravitons.  Each operators results to a butterfly velocity which is given 
in terms of its dimension. In other words, when we are perturbing the bulk metric
by injecting energy, the boundary values of the metric get changed that would excite both operators 
on the boundary.  This is the reason that we get two butterfly velocities in this case. 
Of course at the critical point where both operators have the same dimension we get one butterfly 
velocity. We will make  this point more precise in the next section  where we are considering 
the three dimensional TMG theory.
 
To explore the role of boundary excitations, it is  illustrating to study  $D\geq 5$ dimensional gravities
modified by  Gauss-Bonnet terms. These models have been studied in \cite{Roberts:2014isa} and in 
what follows we use the results of this paper and study butterfly velocity at a distinguished 
point. Let us consider the following action
\be\label{GBD5}
I=\frac{1}{\kappa}\int d^{D}x\,\sqrt{-g}\left[R+\frac{(D-1)(D-2)}{\ell_0^2}
%+\tilde{\alpha} R^{2}+\tilde{\beta} R^{\mu\nu}R_{\mu\nu}
+\gamma\left(R^{\mu\nu\rho\sigma}R_{\mu\nu\rho\sigma}-4R^{\mu\nu}R_{\mu\nu}+R^{2}\right)
\right].
\ee
The butterfly velocity for this model is\cite{Roberts:2014isa}
\be
v_B=\sqrt{\frac{1+\sqrt{1-4\lambda_{GB}}}{2}}\;\sqrt{\frac{D-1}{2(D-2)}},
\ee
where 
\be
\lambda_{GB}=\frac{(D-3)(D-4)}{\ell_0^2}\;\gamma.
\ee
Note that  although the action contains higher derivative terms, the resultant equations of 
motion are still  second order and therefore we find only one butterfly velocity. 
We note, however, that although the equations are second order, the model in general could have two AdS vacuum solutions whose radius are given by the following algebraic equation
\be
\ell^4-\ell^2\ell_0^2+\lambda_{GB}\ell_0^4=0.
\ee
Interestingly enough, when the above equation degenerates  the model
does not have local propagating graviton. This occurs  at $4\lambda_{GB}=1$\cite{Fan:2016zfs}.
In this case the butterfly velocity reads
\be
v_B=\sqrt{\frac{1}{2}}\;\sqrt{\frac{D-1}{2(D-2)}}.
\ee
It is important to note that in this case although the model does not have propagating gravitons,
due to boundary gravitons the butterfly velocity is non-zero. This is indeed very similar 
to what happens in three dimensional Einstein gravity.
%%%%%%%%%%%%%%%%%%%%%%%%%%%%%%%%%%%%%%%%%%%%
%%%%%%%%%%%%%%%%%%%%%%%%%%%%%%%%%%%%%%%%%%%%

\section{Shock waves in 3D TMG model }\label{secII}

In the previous section, we have studied butterfly effete in  $D$-dimensional massive gravities 
that also  includes $D=3$ where we get the New Massive Gravity (NMG) \cite{Bergshoeff:2009aq}. 
In this section we would like to study  butterfly effect for yet another interesting three dimensional
gravity;  Topologically Massive Gravity (TMG) \cite{Li:2008dq}. 
%Due to higher order differential equations, in generic point of the moduli space of parameters of %the model one would expect two butterfly velocities. 

The TMG model is a three dimensional gravity whose action contains the Einstein-Hilbert action
and the  three dimensional gravitational Chern-Simons term
\be
%S_{TMG}&=S_{EH}+S_{CS}\ , \nn\\
S_{TMG}=\frac{1}{16\pi G_N}\left[\int d^3x (R-2\Lambda)+S_{CS}\right]\ ,
\ee
with 
\be 
S_{CS}=\frac{1}{4\mu}\int d^3x\;\epsilon^{\mu\nu\rho}\left[R_{ab\;\mu\nu}\;\omega^{ab}_{\ ,\rho}+\frac{2}{3}\omega^a_{\ b,\mu}\omega^b_{\ c,\nu}\omega^c_{\ a,\rho}\right]\ ,
\ee
where $\omega^a_{\ b,\mu}$ is the spin connection whose inner Lorentz indices
are denoted by $a,b,\cdots$ while the space-time indices are denoted by  $\mu ,\nu ,\cdots$. 

For a generic value of  $\mu$ this model admits an AdS vacuum solution. 
It is conjectured that the TMG model on an asymptotically locally AdS solution with a proper boundary condition would  provide a gravitational dual
for a two dimensional CFT with the following central charges
\be
c_L=\frac{3\ell}{2G_N}\left(1-\frac{1}{\mu \ell}\right),\;\;\;\;\;\;\;\;\;\;
c_R=\frac{3\ell}{2G_N}\left(1+\frac{1}{\mu \ell}\right).
\ee
The model has a critical point at $\mu\ell=1$ where the left central charge vanishes and 
the equations of motion degenerate leading to a log-gravity whose dual theory is a LCFT
\cite{Grumiller:2008qz} (see 
also \cite{{Skenderis:2009nt},{Grumiller:2009mw}}). 

The equations of motion obtained from the above action are
\be
G_{\mu\nu}+\Lambda g_{\mu\nu}+\frac{1}{\mu} C_{\mu\nu}
%+\frac{\gamma}{\mu^2}J_{\mu\nu}+\frac{s}{2m^2}K_{\mu\nu}
=0,
\ee
where
\be
G_{\mu\nu}= R_{\mu\nu}-\frac{1}{2}g_{\mu\nu}R,\;\;\;\;\;\;\;
C_{\mu\nu}= \epsilon\indices{_{\mu}^{\alpha\beta}}\nabla_{\alpha}\left(R_{\mu\nu}-\frac{1}{4}g_{\mu\nu}R\right),
\ee
are  Einstein and the Cotton tensors, respectively.  $AdS_3$ black brane is a solution of the equations of motion,
\be\label{metric1}
ds^2=-\frac{r^2-r^2_h}{\ell^2}dt^2+\frac{\ell^2}{r^2-r^2_h}dr^2+\frac{r^2}{\ell^2}d x^2,\qquad \Lambda=-\frac{1}{ \ell^2}.
\ee

In order to study the shock wave solution in the $AdS_3$ black brane background, one may go through the same procedure considered in the previous section\footnote{The shock wave solution in Mankowski space background for TMG (and NMG) is studied in \cite{Edelstein:2016nml}.}. To do so, one should write the above black brane metric in the Kruskal coordinates,
\be\label{metricKrus}
ds^2=2 A(u v) du\,dv+B(u v)dx^2,
\ee
where
\be
A(u v)=-\frac{2c\ell^2}{(1+c\,uv)^2},\qquad B(uv)=\frac{r_h^2}{\ell^2}\left(\frac{1-c\,uv}{1+c\,uv}\right)^2 .
\ee
Moreover in this case the  tortoise coordinate $r_*$ is
\be\label{tortoisCoord}
r_*(r)=\ell^2\int \frac{dr}{r^2-r_h^2}=\frac{\beta}{4\pi}\left[\log\left(\frac{r-r_h}{r+r_h}\right)-
\log c\right],
\ee
where $c$ is an arbitrary constant. Now, the aim is to study the back-reaction on the metric 
(\ref{metricKrus}) when we inject 
a small amount of  energy towards the horizon so that the equations of motion should be
modified as follows
\be
G_{\mu\nu}+\Lambda g_{\mu\nu}+\frac{1}{\mu} C_{\mu\nu}
%+\frac{\gamma}{\mu^2}J_{\mu\nu}+\frac{s}{2m^2}K_{\mu\nu}
=\kappa T^S_{\mu\nu},
\ee
with $T_{uu}^{S}=\ell E e^{2\pi t_w/\beta}\delta(u)\delta(x)$. Following the procedure presented in the previous section, we will
consider the following ansatz for the back-reacted metric
\be\label{backreactBG}
ds^2=2A(UV)\,dU\,dV+B(UV) dx^2-2A(UV)h(x)\delta(U)dU^2,
\ee
where $U\hspace{-1mm}=u,\hspace{.2cm}V\hspace{-1mm}=v+h(x)\Theta(u)$. Plugging this 
anstaz into the modified equations of motion, one arrives at 
\be
\left(\partial_x +\mu \frac{r_h}{\ell}\right)\left(\partial^2_x-\frac{r_h^2}{\ell^2}\right)h(x)=
-\frac{r_h^3\mu}{2c\,\ell^5}\left[\kappa\,\ell\,E e^{2\pi t_w/\beta}\right]\delta(x),
\ee
which can be decomposed into two differential equations as follows
\begin{numcases}
\displaystyle q'(x)+{a}_1q(x)= \eta\,\delta(x)\label{MMGeq01}\\
\displaystyle h''(x)-a_2^2\,h(x)=q(x),\label{MMGeq02}
\end{numcases}
where 
\be\label{coeffOfsoluMMG}
{a}_1=\frac{r_h\mu}{\ell},\qquad a_2=\frac{r_h}{\ell^2},\;\;\;\;\;\;\;\;
\eta=-\frac{r_h^3\mu}{2{c}\,\ell^5}\left[\kappa\,\ell\,E e^{2\pi t_w/\beta}\right].
\ee
The  equation \eqref{MMGeq01} is indeed Green's function equation whose solution for $x>0$ that falls off at infinity is\footnote{There could be an extra constant in the solution, though it does not change the results and therefore we have set is to zero.}
\be\label{GreenTMG}
q(x)=\eta\,\Theta(x)\;e^{-{a}_1x}.
%\quad \text{or}\quad q(x)=(-\eta\,\Theta(-x)+c_1)\exp\left[-{a}_1\,x\right]
\ee
Treating the function  $q(x)$  as a source for function $h(x)$  and   taking 
into account that the Green's function of the eqaution \eqref{MMGeq02} is given by \eqref{GreenTMG} 
one arrives at
\be
h(x)=-\frac{\eta}{2a_2}\int_{-\infty}^{\infty}\,dy\, \Theta(y)\;e^{-{a}_1\,y-a_2|x-y|}.
%\!\!&\!\!=\!\!&\!\!-\frac{1}{2a_2}\left[e^{-a_2\,x}\int_{-\infty}^{x}\,dy\, (\eta\,\Theta(y)+c_1)
%e^{-({a}_1-a_2)\,y}
%+e^{a_2\,x} \int_{x}^{\infty}\,dy\, (\eta\,\Theta(y)+c_1)e^{-({a}_1+a_2)\,y}\right]\nn
\ee
It is then straightforward to perform the integral. Indeed   for $a_1\neq a_2$ one gets
\bea
h(x)&=&-\frac{\eta}{2a_2}\left[ e^{-a_2\,x}\int_{0}^{x}\,dy\;e^{-({a}_1-a_2)\,y}
+e^{a_2\,x} \int_{x}^{\infty}\,dy \,e^{-({a}_1+a_2)\,y} \right]\cr
&&\cr
&=& -\frac{\eta}{2a_2}\left[\frac{e^{-a_2 x}}{{a}_1-a_2}-\frac{2a_2 \,e^{-{a}_1 x}}{{a}_1^2-
a_2^2}\right],\eea
%while for $x<0$ one finds\footnote{ Since the model, unlike NMG, is not invariant under parity
%we wouldn't expect to get the same shock wave for $x>0$ and $x<0$.} 
%\be
%h(x)=-\frac{\eta}{2a_2}\;e^{a_2\,x} \int_{0}^{\infty}\,dy \,e^{-({a}_1+a_2)\,y} =
%-\frac{\eta}{2a_2}\left(\frac{1}{{a}_1+a_2}\right)e^{a_2\,x}.
%\ee
while  at the special case of ${a}_1=a_2$,  which corresponds to the critical point of the model,   one 
gets
\be\label{logTMG}
h(x)=-\frac{\eta}{2{a_{2}}}\left(x+\frac{1}{2a_2}\right)e^{-a_2 x},
\ee
indicating that the logarithmic mode appears in the spectrum of the model.

Using the explicit expressions for the parameters $\eta, a_1, a_2$ and with a proper choice of $c$, 
one can read the scrambling time and butterfly velocities as follows
\be\label{BFTMG}
t_*=-\frac{\beta}{2\pi}\log\frac{\kappa}{ \ell },\;\;\;\;\;v_B^{(1)}=\frac{2\pi}{\beta a_2}=1,\qquad 
\hat{v}_B^{(2)}
=\frac{2\pi}{\beta a_1}=\frac{1}{\mu \ell }.
\ee
One observes that due to higher derivative terms in the equations of motion, there are two butterfly 
velocities for left moving sector. On the other hand, at the critical point where $a_1=a_2$ the 
dimensions of both operators become  the same resulting to  one butterfly velocity, $v_B^{(1)}=1$.
 
As we have already mentioned in the previous section, the butterfly velocities are given by the 
dimension of operators sourced by metric. 
%When the Einstein equations contains more that  two 
%derivatives, the boundary values of the metric  generically provide sources for different 
%operators with different dimensions resulting to different butterfly velocities. 
To explore this point 
better let us  consider butterfly effect from the dual  2D CFT.

To proceed, we recall  that to diagnose quantum chaos it is useful to study 
out-of-time order four-point correlation function between pairs of local operators
\be
\langle W(t)  V W(t) V\rangle_{\beta},
\ee   
which should be thought of as averaging in the thermal state $| {\beta}\rangle$. In the
present case in order to compute 
this correlation function one  may take advantage of 2D CFT to map the above four-point correlation 
function to a four-point function in a vacuum state. More precisely, using the transformation
\be\label{CFTtransfromedCoord}
z(x,t)=e^{\frac{2\pi}{\beta}(x+t)},\qquad \bar{z}(x,t)=e^{\frac{2\pi}{\beta}(x-t)},
\ee
one leads to compute the four-point function  $\langle W(z_{1},\bar{z}_{1}) V(z_{2},
\bar{z}_{2})W(z_{3},\bar{z}_{3}) V(z_{4},\bar{z}_{4})\rangle _{vac}$. Actually by making use of 
2D conformal symmetry one has 
\be\label{Euclidean4p}
\frac{\langle W(z_1,\bar{z}_1) V(z_2,\bar{z}_2) W(z_3,\bar{z}_3) V(z_4,\bar{z}_4) \rangle}{\langle 
W(z_1,\bar{z}_1) W(z_3,\bar{z}_3)\rangle \langle V(z_2,\bar{z}_2) V(z_4,\bar{z}_4) \rangle}
 = f(z,\bar{z}),
\ee
where $f(z,\bar{z})$ is an  arbitrary function and 
\be\label{CRat}
z=\frac{z_{12}z_{34}}{z_{13}z_{24}},\qquad \bar{z}=\frac{\bar{z}_{12}\bar{z}_{34}}
{\bar{z}_{13}\bar{z}_{24}}.
\ee 
There is a well-known procedure to compute the function $f(z,\bar{z})$ (see {\it e.g.} 
\cite{Cornalba:2006xm}). In fact in our case  the result is\cite{Perlmutter:2016pkf}
\be\label{CB0}
f(z,\bar{z})=2\pi i\sum_{\mathcal{O}(\Delta,s)}\alpha^2_{\mathcal{O}}\, \frac{\Gamma{(\Delta+s)}
\Gamma{(\Delta+s-1)}}{\Gamma^{4}(\frac{\Delta+s}{2})} z^{1-s} \eta^{\frac{\Delta-s}{2}},
\ee
where the sum runs over the conformal primary operators ${\cal O}$ whose dimension and
spin are given by   $\Delta$ and $s$, respectively and, $\eta=\frac{\bar{z}}{z}$. Moreover, $\alpha^2_{\mathcal{O}}= \alpha_{WW\mathcal{O}}\alpha_{VV\mathcal{O}}$, with {\it e.g.} $\alpha_{WW\mathcal{O}}$ is the OPE coefficient in $WW$ operator product. By making use 
of the definition of cross ratios \eqref{CRat} and with the desired time-ordering that
fixes the expressions of $z$ and $\bar{z}$\cite{Perlmutter:2016pkf}, 
one finds
\be
f(z,\bar{z})\approx 2\pi i\sum_{\mathcal{O}(\Delta,s)}\;\;\;\;\;\;\;\hspace{-0.8cm}
\alpha^2_{\mathcal{O}}\, 
\frac{\Gamma{(\Delta+s)}\Gamma{(\Delta+s-1)}}{\Gamma^{4}(\frac{\Delta+s}{2})(-\epsilon_{12}^*
\epsilon_{34})^{s-1}} e^{\frac{2\pi}{\beta}(s-1)[t-\frac{\Delta-1}{s-1}x]},\nn
\ee
where  $\epsilon_{ij}=i(e^{\frac{2\pi}{\beta}i\epsilon_i}-e^{\frac{2\pi}{\beta}i\epsilon_j})$ with 
$i\epsilon_i$ being an infinitesimal Euclidean time associated to each of four 
operators\footnote
{Note that in order to get the right thermal averaging four point correlation one should  choose
 $\epsilon_1<\epsilon_2<\epsilon_3<\epsilon_4$.}. Using this expression, one can read 
 the Lyapunov exponent and butterfly velocity as follows (see also \cite{Roberts:2014ifa})
 \be\label{BFsd}
\lambda_{L}=\frac{2\pi}{\beta}(s-1),\;\;\;\;\;\;\;\;\;\;\;v_B(\Delta,s)=\frac{s-1}{\Delta-1}.
\ee
It was shown in \cite{Perlmutter:2016pkf} that for a CFT whose gravitational dual is provided by
Einstein gravity  the main contribution to $f(z,\bar{z})$ comes from spin-2 operator of 
the lowest twist that is energy-momentum tensor. On the other hand since in the present case where
the metric sources two operators whose dimensions can be taken as closed as we want
by tuning the parameters of the model, one would expect that the operator which is dual to
the massive spin-2 mode should also contribute to the four-point function. 
Therefore we arrives at two butterfly velocities which are given in terms  of spin and 
dimension of operators as \eqref{BFsd}.

Let us now apply this result to our cases. For TMG model where the spectrum 
contains massive and massless gravitons one gets two spin-2 operators with 
dimensions $\Delta^{(1)}=2$ and $\Delta^{(2)}=1+\sqrt{1+\ell^2 M^2}$ with $M^2=\mu^2\ell^2-1$.
Plugging these expressions into the equation \eqref{BFsd} one arrives at
\be
v_B(\Delta^{(1)},2)=1,\;\;\;\;\;\;\;\;v_B(\Delta^{(2)},2)=\frac{1}{\mu\ell},
\ee
in agreement with  \eqref{BFTMG}. On the other hand for the NMG model one gets
\be
v_B(\Delta^{(1)},2)=1,\;\;\;\;\;\;\;\;v_B(\Delta^{(2)},2)=\frac{1}{\sqrt{1+\ell^2 M^2}},
\ee
which is the same as that we have found in the previous section for $D=3$.

We have also seen that at the critical point where the massive spin-2 degenerate
with the massless graviton leading to the log-gravity, two butterfly velocities
coincide. It is also illustrative to see this effect from proper conformal block decomposition 
approach. Actually 
%at the critical point, the dual operators to bulk massive and massless spin-2 fields degenerate.  
in this case, the  conformal block decomposition \eqref{CB0} should be  substituted 
 with\cite{Hogervorst:2016itc}\footnote{ Note that here one has
 $\alpha^2_{a b} = \alpha_{WW a}\hspace{1mm}\alpha_{VV b}$\cite{Hogervorst:2016itc}.}
  \be
f(z,\bar{z}) = 2\pi i\left(\alpha^2_{T \tau}+\alpha^2_{\tau T}+\alpha^2_{\tau\tau}\frac{\partial}{\partial\Delta}\right)\frac{\Gamma{(\Delta+2)}\Gamma{(\Delta+1)}}{\Gamma^{4}(\frac{\Delta+2}{2})}\hspace{1mm} z^{-1} \eta^{\frac{\Delta-2}{2}}\bigg{|}_{\Delta \rightarrow 2},
\ee
where $T_{\mu\nu}$ and $\tau_{\mu\nu}$ are  the energy-momentum tensor and its logarithmic counterpart, respectively.  By making use of the proper cross ratios one arrives at 
\be
f(t,x)\approx 2\pi i\hspace{1mm}  \frac{4(3 \alpha^{2}_{T \tau}+3 \alpha^{2}_{\tau T}+4 \alpha^{2}_{\tau \tau})}{(-\epsilon_{12}^*
	\epsilon_{34})}\left[1-\frac{3\alpha^{2}_{\tau\tau}}{(3 \alpha^{2}_{T \tau}+3 \alpha^{2}_{\tau T}+4 \alpha^{2}_{\tau \tau})} 2\pi x/\beta\right]\hspace{1mm}e^{\frac{2\pi}{\beta}(t-x)
},
\ee
which has the same structure as that of  (\ref{logNMG}) and (\ref{logTMG})\footnote{The logarithmic shock wave solution just for left moving sector in TMG might be understood from the non-parity invariant structure of OPE coefficients in this theory.}. 
More precisely to reproduce these equations one should set 
\bea
&&\text{NMG}:\hspace{1cm}\alpha^{2}_{T\tau}+\alpha^{2}_{\tau T}= -\frac{7}{3}\alpha^{2}_{\tau\tau} \cr\nonumber\\
&&\vspace{-1cm}\text{TMG}:\hspace{1cm}\alpha^{2}_{T\tau}+\alpha^{2}_{\tau T}= -\frac{11}{6}\alpha^{2}_{\tau\tau}.
\eea
Note also that we have one butterfly velocity $v_B=1$ as expected.

\section{Conclusions}
 
In this paper, we have studied butterfly effect in $D$-dimensional gravitational
theory containing higher order derivatives. The higher order terms consist of 
Ricci scalar and Ricci tensor squared. For generic values of the parameters of the model
we have found two butterfly velocities, though at the critical points where the equations 
of motion degenerate these two velocities coincide. The observation of our paper may be
explored as  follows.  

From holographic renormalization
 \cite{deHaro:2000xn} in the context of gauge/gravity duality  we know that the boundary value
 of a bulk field (non-normalizable mode) should be identified with the source of the dual 
 operator whose dimension and spin are fixed by the mass and the spin of the bulk field.
 In particular, for Einstein gravity the metric is dual to the energy-momentum tensor of the 
 boundary theory. 
 
 Going to higher derivative gravities, typically the corresponding equations of motion
consist of higher order differential equations so that the metric may be fixed by given 
several boundary conditions. The boundary condition (if corresponds to non-normalizable mode)
might be identified with sources of dual operators all of which have spin-2, though their 
dimensions would be different.

In particular for the models we have considered in this paper, the excitation of the metric 
contains  massive and massless gravitons so that the dual theory should have two 
spin-2 operators. When we are perturbing the bulk geometry, the boundary values of 
metric would also exciting the corresponding boundary operators. To each 
spin-2 operators, one may associate a butterfly velocity which is determined by the
dimension of the corresponding operator\footnote{Note that  scalar field or vector field 
cannot lead to butterfly velocity.}
 
Actually  this observation should be thought of as a generalization of the results 
presented in  that \cite{Perlmutter:2016pkf} where it was shown that in any 
holographic CFT whose gravitational dual is provided by Einstein gravity the butterfly 
velocity is determined by  the energy-momentum tensor (see also \cite{Roberts:2014ifa}). 

To explore the role of the boundary value of the metric, we have also studied butterfly effect
in $D$-dimensional gravitational theories corrected by Gauss-Bonnet term. In this case since the 
equations of motion are still second order one gets one butterfly velocity. There is, however,
a point in the moduli space of the parameters of the model where the model does not have 
propagating gravitons on the bulk, though it still has boundary gravitons. In this case
we still have non-zero butterfly velocity showing the importance of the boundary modes.
Actually the situation is very similar to that of three dimensional gravity and indeed 
it can be seen that at this point the action reduces  to five dimensional gravitational 
Chern-Simons action.

It is also interesting to compute butterfly velocity for a gravitational theory whose spectrum  
contains only  a massive graviton (no massless graviton).  Let us  consider the following
 particular model
(see  \cite{Hassan:2011vm})
\be
S=\int d^4x \sqrt{-g}\left[\frac{1}{2\kappa^2}\left(R+\frac{6}{L^2}-\alpha^2\left[(\Tr\KK)^2-\Tr\KK^2\right] \right)-\frac{1}{4e^2}F_{\mu\nu}F^{\mu\nu}\right],
\ee
where $\mathcal{K}\indices{^\mu_\nu}=\sqrt{g^{\mu\lambda}f_{\lambda\nu}}$. It is straightforward
to write the equations of motion of the above action. Then setting 
 $f_{\mu\nu}=\text{diag}(0,0,1,1)$ and for 
 non-zero component of gauge field $A_0=a(r)$, one finds the following black hole solution
 \cite{{Blake:2013bqa},{Vegh:2013sk}}, 
\be
ds^2=\frac{L^2}{r^2}\left(-b(r)\,dt^2+\frac{1}{b(r)}dr^2+dx^2+dy^2\right),\;\;\;\;
A_t= \mu\left(1-\frac{r}{r_h}\right),\label{massiveMetric}
\ee
where
$b(r)= 1-\alpha^2 r^2-M r^3+\frac{\mu^2\,r^4}{\gamma^2 r_h^2}$, with $\gamma^2=
\frac{2e^2 L^2}{\kappa^2}.$
Going through the procedure we presented in the previous sections one can find the 
butterfly velocity as follows
\be\label{B}
v_B=\frac{1}{2} \sqrt{3-\alpha^2 r_h^2  -\frac{\mu^2r_h^2}{\gamma^2}}.
\ee
This should be compared with that of Einstein gravity which is given by setting $\alpha=0$.

Of course in this paper we have just considered  cases where the bulk equations of motion
are at most fourth order and therefore we have obtained two butterfly velocities. Going beyond 
fourth order we may get more velocities.

%%%%%%%%%%%%%%%%%%%%%%%%%%%%%%%%%%%%%%%%%%%%%%%%%
\section*{Acknowledgments}

We would like to thank Amin Akhavan, Amin F. Astaneh, Ali Mollabashi, Farzad Omidi, 
Mohammad M. Mozaffar, Ahmad Shirzad and Mohammad R. Tanhayi for useful discussions.
We also acknowledge the use of M. Headrick's excellent Mathematica package ``diffgeo''. 
We would like to thank him for his generosity. This work is supported by Iran National Science Foundation (INSF).

\end{document}